# Oriented Convex Containers of Polygons - II

R Nandakumar

- Amrita School of Arts and Sciences,
Idapalli North, Kochi 682024, India.
(nandacumar@gmail.com)

**Introduction:** This article extends the earlier article on 'Oriented Convex Containers' [1]. Let us recall the definition of an oriented convex region as a convex region with a direction of symmetry. The above article had touched upon isosceles triangles, rectangles and ellipses. A question raised in [1] was: given a triangle T, characterize the minimum area and minimum perimeter isosceles triangles that contain T. Recently, major progress has been achieved on this question in [2]. In this document, we examine some more possible oriented containers and raise some questions.

**Question 1:** How does one find the smallest semicircular region ('semidisk' hereafter) that contains a given set of points on the Euclidean plane?

**Observation:** it is sufficient to form the convex hull of the given set of points and to contain this hull; this is a feature common to all optimal containment problems we discuss here. Let there be N vertices on this convex hull. Hereafter we refer to this convex hull simply as hull.

**Lemma 1:** The diameter of the smallest semidisk cannot be empty of the points to be contained.

**Proof:** Indeed, if a candidate smallest semidisk containing a set of points is such that its diameter does not pass through any of the points (some points may lie on the arc but that is immaterial), by shifting the candidate smallest semidisk perpendicular to this diameter a bit, we can have it to contain all the points but NOT passing through any of them – and such a semidisk is clearly suboptimal.

**Lemma 2:** The diameter of the smallest semidisk either has one vertex of the hull at its midpoint or it contains exactly one edge of the convex hull and that edge of the hull contains the midpoint of the diameter.

**Proof:** The diameter of the smallest semidisk cannot contain more than one edge of the hull (due to convexity).

Assume the diameter of the smallest semidisk contains a single vertex of the hull (call this vertex V) that is not coincident with the midpoint of the diameter. Now, consider rotating the semidisk slightly about the normal to the plane passing thru the diameter midpoint. Since the candidate container is a semidisk, none of the vertices of the hull initially on or within the arc will leave the semidisk and V would have moved into its interior, leaving the diameter devoid of hull vertices. That shows the candidate semidisk to be suboptimal, as argued in proof of lemma 1.

By the same rotation, we can deal with the case where the diameter of the candidate smallest semidisk contains an edge of the hull that in turn **does not** contain the midpoint of the diameter in just the same manner as above. See figure below. It shows the case where the candidate semicircle contains an edge of the hull which in turn does not contain midpoint of the diameter.

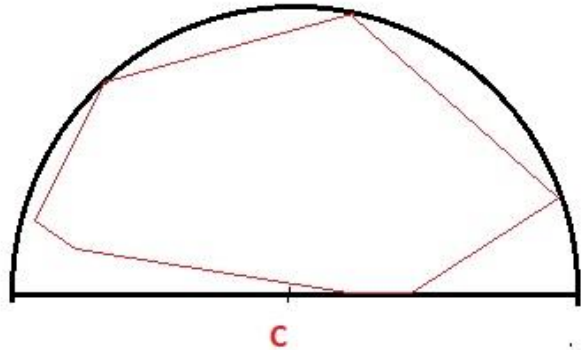

c

On the basis of these lemmas, we now formulate an algorithm for finding the smallest semidisk containing the N vertex convex hull of the input points.

**Algorithm 1:**

With diameter fixed along each edge E of the hull, we form the smallest semidisk that contains the entire hull. Then select that edge that gives the smallest semidisk and return it.

To find the smallest semidisk, we need to do the following:  Given a line segment (edge of hull ) E, find that point P on it such that P is the center of the smallest semidisk with E as diameter and contains the hull.

> Let A and B be the end points of E.  Let d be the distance from A to B. Let x indicate distance measured from A to B along E. So x varies from 0 to d.

- For every vertex V on the hull other than A and B, plot against x, a graph of the distance from V to a point on E at a distance x from A. This plot will in general be a continuous curve above x. If the perpendicular from V to line determined by E does not pass thru segment E, then,  the curve will be monotonically increasing or decreasing as x goes from 0 to d; otherwise, it will have a single minimum at an intermediate value and maxima at x =0 and d. Since V runs through N-2 vertices, we have N-2 such curves lying above the x interval (0 - d).

- From the set of all points that lie on **all** these N-2 curves, for each value of x, select that point with highest y coordinate. This will result in a continuous curve formed out of segments from several of the N-2 curves we have plotted. Call this curve the top envelope of all the N-2 curves.

> Find where the top envelope has the minimum value. That value of x gives a point on edge E that is at the **least maximum distance** from all other points on the hull. This point is the center of the smallest halfdisk with edge E contained in its diameter.

> Since we have to find intersections among N cuves to find the top envelope, the processing for a single edge E will have $O(N^2)$ complexity.

We could repeat the above process with each of the N edges of the hull as E and select the smallest from the N resulting semidisks, thus resulting in an O(N³) algorithm. But we make the following crucial observation: for a given edge $E_i$, if $V_j$ is the vertex that is at the largest min distance from $E_i$, then, if we move counterclockwise from edge $E_i$ to $E_{i+1}$, then the vertex that is at largest min distance from $E_{i+1}$ is either $V_j$ itself or a vertex $V_{j+k}$ that comes after $V_j$ as we proceed CCW from $V_j$ along the boundary. Moreover, no vertex between $V_j$ and $V_{j+k}$ will be the maximizer of the minimum distance from any edge – so, if we progress sequentially along the edges, the progress along the vertices will also be non-reversing.

The above observation limits the number of vertices examined to find the farthest from each successive edge to a constant number on the average – in other words, we need to examine only a total of O(N) such vertex-edge pairs after the O(N²) processing for the edge with which we begin processing. This property is essentially what drives the rotating calipers technique [3] and reduces the complexity of the algorithm to O(N²)

Perhaps a smart linear programming approach could achieve a further reduction of complexity (a linear time one exists for the smallest containing circle of a system of points. For more information see [4]).

**Remark:** It is easily seen that for some point sets, the smallest containing circle has less area than the smallest containing semicircle (for example, if the hull is close in shape to a circle) and vice versa (for example, if the hull is close to a semicircle). So, intuitively, if we slowly expand the hull to be contained from a semidisk to a full disk, at some intermediate stage of the evolution, the area of the smallest containing semicircle will be equal to the smallest containing circle.

## Further questions

**Definition:** A circular segment ('segment' hereafter) is a part of a circular region cut out by a chord. To distinguish which of the two region into which a chord divides a disk, one has to specify the angle subtended at the center by the arc bounding the segment. For the larger segment the angle measure will be greater than $\pi$.

**Question 2: How does one find the smallest circular segment that contains a given set of points?**

Note that in the case of smallest semicircle, one needs only to find center and radius. For a segment, we need to find center, radius and also the angle subtended at the center by the arc of the circular segment. Moreover, the word 'smallest' has a unique meaning for the case of smallest semicircle. In the case of circular segments, there are two variants for a given hull being contained – the circular segment with smallest area (we call this $S_A$) and with smallest perimeter (call this $S_P$).

**Claim:** Lemma 1 and 2 apply to both smallest segments – $S_A$ as well as $S_P$ - with small modifications. Ie. The smallest segment containing the hull is such that the chord bounding the segment contains an edge of the hull and this edge contains the midpoint of this chord.

**Proof:** we reuse the proofs of lemma 1 (for candidate segments with angle measure less than pi, the argument is as before before - slide the hull in a direction perpendicular to the chord; for candidate segments with angle measure greater than pi, the validity of lemma 1 is immediately obvious) and of lemma 2 (basically same argument as before except that the center of rotation is not the midpoint of the chord but the center of the full circle of which the candidate segment is a part).

**Note:** Although both smallest containing segments for a given hull have the property that the chord bounding the segment contains an edge of the hull, there is a possibility that for some hulls, $S_A$ and $S_P$ *might be* based on different edges of the hull and their radii and angle measures could be different. We do not yet have a concrete example of such a hull.

Here, we suggest a broad approach to find both $S_A$ and $S_P$ for any given hull. There appear to be two cases: (1) the angle measure of the smallest segment is greater than $\pi$ and (2) less than $\pi$. If the center of the smallest disk containing the hull is within the hull, we have case 1, else case 2.

**Case 1 (Guess):** For this case, both $S_A$ and $S_P$ can be obtained by first finding the smallest disk within which the hull is inscribed and cutting off from the disk a suitable circular segment that lies outside the hull. It appears that in this case, both $S_A$ and $S_P$ are the same.

**Case 2 (Guess):** As described in algorithm 1, we can, for each edge $E_i$ of the hull, construct the smallest semicircle S with midpoint on this edge and containing all other vertices (as described above) and then 'shrink' this semicircle in area (or perimeter ) into a segment that is part of a circle with greater radius than S but with angle measure less than $\pi$ and still containing the entire hull; then we can select the overall smallest segment from n candidates. We do not yet have an efficient algorithm that does this shrinking.

**Lemma 3:** For a given hull, consider the segments of minimum area and perimeter found with a particular edge of the hull E determining the chord. Although the chords of the two segments overlap, the midpoints of the chords *need not* coincide – the axes of symmetry of the two segments are then parallel, not coincident.

**Proof:** We show an example below:

Consider an isosceles right triangle. For it both $S_A$ and $S_P$ are the same – a semicircle (call it S) in which the right triangle is inscribed. Now, consider altering the triangle continuously by moving the apex along the boundary of S as shown by the arrow – we pass through a sequence of right triangles which get progressively thinner – and finding the two smallest containing segments with chord coincident with the hypotenuse of the triangle.

For some finite distance moved by the apex, the semicircle is still the best answer for both types of containing segments for the new right triangle. But eventually, the smallest segment with its chord lying along the hypotenuse of the triangle and containing the new thin right triangle will become one with larger radius than S but thinner than a semicircle – and the center of the circle determining this segment will have shifted from O into the 4th quadrant as shown. As can be verified numerically, for the two kinds of smallest containing segments, the shifting of the center happens at different stages of the motion of the apex and at different rates - indeed, for the least perimeter segment, shift begins later and is slower - and so the axes of symmetry of the two optimal containing segments with the hypotenuse of the triangle as chord will only be parallel and not coincident.

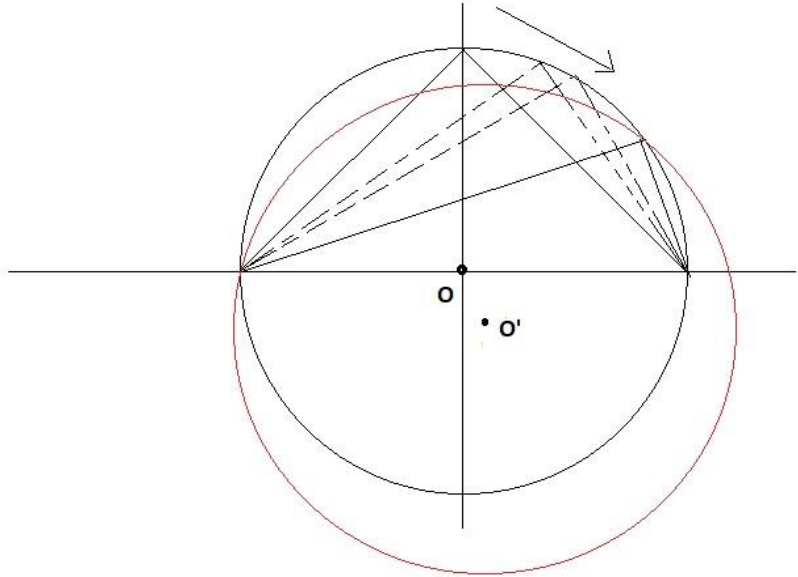

**Question 3:** Are there convex regions for which $S_A$ and $S_P$ have their chords inclined by any finite angle to each other? If so for which convex region are the chords of the two containers maximally apart in direction?

We conclude this note by mentioning another oriented convex container:

**Question 4:** How does one find the smallest **sector** that contains a given set of points (again, there are two different variants – area and perimeter)? Can optimal algorithms be formulated beginning with the smallest containing semicircle?

**Acknowledgement:** Thanks to K Sheshadri for several helpful discussions.

**References:**

1. R Nandakumar – 'Oriented Convex Containers of Polygons' - https://arxiv.org/abs/1802.10447
2. Gergely Kiss, Janos Pach, Gabor Somlai - 'Minimum Area Isoscles Containers' - https://arxiv.org/abs/2001.09525
3. Rotating Calipers Method: https://en.wikipedia.org/wiki/Rotating_calipers
4. David Mount – 'Lecture Notes in Computational Geometry' - http://www.cs.umd.edu/~mount/754/Lects/754lects.pdf